\documentclass[manuscript]{acmart}
\usepackage{booktabs}
\usepackage[normalem]{ulem}
\usepackage{mathtools}

\setcopyright{none}
\renewcommand\footnotetextcopyrightpermission[1]{}
\DeclarePairedDelimiterX{\infdivx}[2]{}{}{%
  #1\;\delimsize\|\;#2%
}
\newcommand{\infdiv}{\infdivx}

\def\BibTeX{{\rm B\kern-.05em{\sc i\kern-.025em b}\kern-.08emT\kern-.1667em\lower.7ex\hbox{E}\kern-.125emX}}

\setcopyright{acmcopyright}
\acmJournal{PACMHCI}
\acmYear{2020} \acmVolume{4} \acmNumber{GROUP} \acmArticle{12} \acmMonth{1} \acmPrice{15.00}\acmDOI{10.1145/3375192}

\begin{document}

%
\title{Measuring the Diversity of Facebook Reactions to Research}

%

%
\author{Cole Freeman}
\email{cole.freeman9@gmail.com}
\orcid{1234-5678-9012}
\affiliation{%
  \institution{Northern Illinois University}
  \streetaddress{1425 Lincoln Hwy}
  \city{DeKalb}
  \state{Illinois}
  \postcode{60115}
  \country{USA}
}

\author{Hamed Alhoori}
\affiliation{%
  \institution{Northern Illinois University}
  \streetaddress{1425 Lincoln Hwy}
  \city{DeKalb}
  \state{Illinois}
  \postcode{60115}
  \country{USA}
}
\email{alhoori@niu.edu}

\author{Murtuza Shahzad}
\affiliation{%
  \institution{Northern Illinois University}
  \streetaddress{1425 Lincoln Hwy}
  \city{DeKalb}
  \state{Illinois}
  \postcode{60115}
  \country{USA}
}

%
\renewcommand{\shortauthors}{Cole Freeman, Hamed Alhoori, \& Murtuza Shahzad}
%
\begin{abstract}

Online and in the real world, communities are bonded together by emotional consensus around core issues. Emotional responses to scientific findings often play a pivotal role in these core issues. When there is too much diversity of opinion on topics of science, emotions flare up and give rise to conflict. This conflict threatens positive outcomes for research. Emotions have the power to shape how people process new information. They can color the public's understanding of science, motivate policy positions, even change lives. And yet little work has been done to evaluate the public's emotional response to science using quantitative methods. In this paper, we use a dataset of responses to scholarly articles on Facebook to analyze the dynamics of emotional valence, intensity, and diversity. We present a novel way of weighting click-based reactions that increases their comprehensibility, and use these weighted reactions to develop new metrics of aggregate emotional responses. We use our metrics along with LDA topic models and statistical testing to investigate how users' emotional responses differ from one scientific topic to another. We find that research articles related to gender, genetics, or agricultural/environmental sciences elicit significantly different emotional responses from users than other research topics. We also find that there is generally a positive response to scientific research on Facebook, and that articles generating a positive emotional response are more likely to be widely shared---a conclusion that contradicts previous studies of other social media platforms.

\end{abstract}

%
%

%

%
\maketitle

\section{Introduction}

The rate at which information is being shared and reacted to on social-media platforms is increasing year by year. In June 2019, 293,000 new posts and 510,000 new comments were made on Facebook every minute~\cite{noyes}. These platforms have further accelerated the speed of content evaluation and feedback with features such as click-based reactions. These relatively new features provide users with a quick and easy way to respond to content in a manner that is still personal and expressive. They have been introduced on platforms such as Facebook, which expanded its reaction palette in 2016, and LinkedIn, which increased the number of click-based reactions available to users in 2019. Until now, however, very little research has been undertaken to advance our understanding of these novel features. 

Click-based reactions offer obvious benefits to researchers in the fields of bibliometrics and alternative metrics (or altmetrics)---a growing area of interest that takes into consideration the dissemination of a research outcome via multiple social media platforms ~\cite{costas:10.1002/asi.23309, Sugimoto:10.1002/asi.23833, THELWALL2018237,Didegah2018}. Previous studies have used citations as the gold standard for understanding and predicting the influence that scholarly research has on the scientific community itself ~\cite{alhoori2019anatomy}, but evaluating the emotional impact that work may have on society has remained largely untouched. Click-based reactions may provide a way to approach this question of how research is affecting society.

It is often the case that advancements in scientific understanding are preceded by advancements in measurement. Though there are well-established methods and tools for analyzing the sentiment of texts, the novelty of click-based reactions has meant that there are not the same resources available for analyzing emotions expressed through clicks. In this paper we present a heuristic for analyzing click-based responses to social-media content. We borrow from psychological research the concepts of emotional valence, intensity, and diversity, using them to gain an improved understanding of how Facebook users are responding to scientific research through their reactions. We develop metrics for these emotional concepts after careful analysis of a dataset of Facebook reactions that we collected for this study. Finally, we apply our newly developed methods and metrics to our data. We are most interested in learning about aggregate behavior and sentiment toward topics of science rather than toward individual articles, therefore we cluster the articles in our dataset using an LDA topic model and perform statistical testing to learn how user expressed sentiment changes based on the shared content.

\section{Background}

\begin{figure}
  \includegraphics[width=0.98\textwidth]{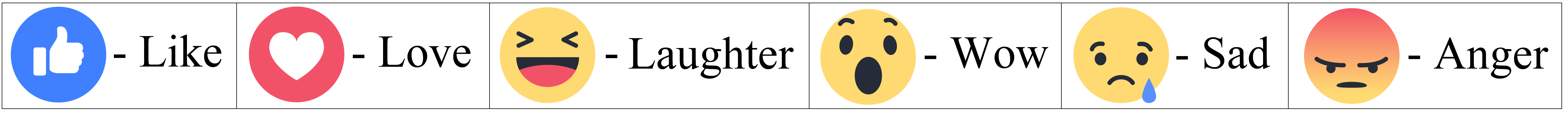}
  \caption{The six click-based reactions available to users on Facebook.}
  \label{fig:emoji_map}
\end{figure}

We begin by defining a few important terms:

\begin{itemize}
    \item \textbf{Click-based reactions}: features on social-media platforms that allow users to leave a quick and easy response to content; click-based reactions are non-textual and are related to textual emojis, which convey emotional responses through pictograms; on Facebook, click-based reactions include the six buttons ``Like,'' ``Love,'' ``Wow,'' ``Laughter,'' ``Sad,'' and ``Anger''; Figure~\ref{fig:emoji_map} shows the six click-based reactions from Facebook. 
    \item \textbf{ Five special reactions}: the five click-based reactions: ``Love,'' ``Wow,'' ``Laughter,'' ``Sad,'' and ``Anger.'' 
    \item \textbf{Page visibility}: the number of followers a Facebook page has; Facebook allows users to like or follow pages; after following a page, a user will begin to see content shared by or on those pages on their own timeline. 
    \item \textbf{Shares}: the number of public Facebook pages an article has been posted onto.
    \item \textbf{Reshares}: the number of times users have re-shared a public post of an article onto another private or public page. Our dataset contains reactions to the initial share and the number of ``Reshares,'' but we do not have reaction data from article ``Reshares.''
    \item \textbf{Articles}: our dataset consists of Facebook reactions to research related articles; for convenience, these are sometimes referred to as ``documents,'' especially in the description of feature transformation.
\end{itemize}

\subsection{Measuring emotion}

When evaluating the emotional responses that a group of users have to social media content, three factors are particularly relevant: the \textit{valence}, \textit{intensity}, and \textit{diversity} of the response. In psychological literature, these factors usually refer to the emotional response of an individual~\cite{Lench2011-tt, Fredrickson2001-ux, Verduyn2013-sx, Compton2003-oz, Ram2011-zb}. Since we are interested in measuring aggregate responses here, we are instead treating these three variables as indications of aggregate response to content. Taking measurements of individual users would not be possible with the data we collected for a number of reasons. First and foremost, we took precautions to avoid collecting information that could be used to identify specific users; measuring more granular emotional responses would necessarily require more information about the individuals. Secondly, each user can only click one reaction on a given post (they can, however, share and provide a reaction to a post). Although this information is useful in finding group responses, it is not at all helpful for identifying when an individual has a diverse or conflicted response to a given paper. 

The valence of response denotes the ``direction'' of the emotion (i.e., positive or negative). Positive emotions promote social connections and playfulness~\cite{Fredrickson2001-ux}, whereas negative emotions promote wariness and avoidance, though they sometimes capture more attention than positive emotions~\cite{Lench2011-tt,Bower2001-zq}. Valence is a binary measurement, and it is important to note that negative responses can have beneficial effects for both individuals and communities~\cite{Bower2001-zq}. The intensity of an emotional response reflects the importance of information to respondents~\cite{Verduyn2013-sx, Verduyn2012-xx, Verduyn2009-ib}. As the intensity of emotion increases, the object of the emotion captures more of individuals' attention~\cite{Compton2003-oz,Lench2011-tt,Levine2009-st,ross2018diffusion}. Intensity can also be used as an indicator of how long information will be retained in a person's memory~\cite{Bower2001-zq}. More intense emotions are also more likely to influence decision making~\cite{Gilbert1998-kn}. 

The final factor we use here is diversity of response. It indicates the degree to which the user response to a post simultaneously point in different emotional directions. It can indicate that content is controversial or that there is not consensus about how a subject should be received or interpreted by a group of responders. A higher diversity of emotional response about a post or article indicates that there is much variation in individuals' responses and that there is not consensus about a particular subject or issue~\cite{Ram2011-zb}. 

\subsection{Topic models}
A topic model is a category of statistical model used in Natural Language Processing (NLP) to discover latent semantic structures or significant word groupings that occur in a corpus of texts. They are often built with a large body of documents as a way to preserve the distinct properties of each document while also giving a brief description of each as a unique mixture of topics. Topic models facilitate useful basic tasks in machine learning, such as summarization of texts, determination of document relevance, detection of novelty, as well as classification. 

Latent Dirichlet allocation (LDA) topic models were first applied in a machine learning context by Blei et al.~\cite{blei2003latent}. The underlying idea of LDA is that documents can be represented as ``random mixtures over latent topics,'' and that each topic can be represented as a probability distribution over words. It is this type of topic model that has been the most widely applied in text mining literature. An LDA model takes as input a corpus of documents and a given number of topics $t$ that the user would like to discover, and defines $t$ groups of words or n-grams that frequently co-occur. Each document can then be represented as a distribution over the $t$ topics. 

\section{Literature review}

Studies in social-media analytics tend to focus either on text, using approaches such as NLP, sentiment analysis, or opinion mining to arrive at and support research conclusions, or on the proliferation of content through online communities~\cite{Hancock2008, Pirzadeh2014, Gabielkov:2016:SCG:2964791.2901462,Siravuri2018}. These approaches have proved effective for understanding or predicting many aspects of human behavior, but they leave a number of other expressive signals unexamined. Click-based reactions, on the other hand, are a relatively underutilized resource in social-media research. Examples of quick-draw, ready-made expressive features are becoming increasingly prevalent across many platforms, and as such have attracted some amount of attention from researchers in the past few years. 

Several studies have been conducted to measure and understand emotions using social media data ~\cite{Giuntini,Graziani,Raad,Badache,shared_feelings}. Tian et al.~\cite{Tian2017-tp} studied the way Facebook users modify the sentiment of their comments with emojis. They targeted posts on public news pages, comparing three channels of expression: natural language, emojis, and reactions. They found that generally the emotional content of emojis and reactions correspond, but that in instances of sarcasm or politeness, the two channels could express different meanings. Krebs et al.~\cite{Krebs2017} used customer satisfaction data gathered from Facebook to train a model using convolutional and recurrent neural networks (CNN and RNN) and predict the reaction distribution for a given post. Basile et al.~\cite{Basile2017-uq} combine NLP and sentiment analysis of Facebook reactions to build a regression model that predicts news controversies in Italian media.

There have been several studies on community building, social interaction, and identity on Facebook and other social-media platforms. Rohde et al.~\cite{Rohde2004} and Hewitt et al.~\cite{hewitt2006crossing} were early examples of how social-media data could be used to study inter-community interaction and identity formation online. Burke et al.~\cite{Burke:2016:OMF:2818048.2835199} found that Facebook users are more likely to respond with greater emotional intensity in both positive and negative directions to community members' posts if their friend networks are smaller and more greatly connected. Thagard and Kroon~\cite{thagard2006emotional} highlighted the role emotional consensus plays in group decision making community cohesion. Their conception of emotional consensus was built on the idea that groups reach accord when all party members communicate the valences they associate with each possibility. The authors argued that consensus is reached in part through rational discussion and argument, but that we overemphasize the importance of this because we tend to believe that humans behave as rational actors. They looked to psychological research to show that group behavior is driven much more by nonverbal forms of communication such as ``facial expressions, voices, postures, [and] movements.'' In working toward consensus, emotionally stronger positions are ``contagious,'' spreading through the group and finally having the greatest influence on outcomes. 

Kumar et al.~\cite{Kumar:2018:CIC:3178876.3186141} look at conflict and confrontations between communities on Reddit. They define communities by the users who participate in distinct forums on the site (i.e., ``subreddits''), each of which caters to specific interests of users and are curated by page moderators. This definition of community as a well-defined space where members frequently visit and interact with other members works well within the structure of Reddit. Reza et al.~\cite{zafarani_abbasi_liu_2014} label this type of community where users understand that they are a member of a community and interact with other members of their community more than nonmembers as \textit{explicit}.

Participation on Facebook can be described as explicit, but the user experience there is not necessarily centered around community pages, but rather around a kind of ``commons'' area---the news feed---where users interact with family and friends and see content that algorithms predict will be of interest to them. Facebook members define their interests and then receive content catered to their wants. The result is that a single user can follow and interact with content on many different pages, often without even directly moving onto the pages from which the content originates. Reza et al.~\cite{zafarani_abbasi_liu_2014} refer to participation in an unacknowledged community as \textit{implicit}.

Some studies have shown that a person's emotions, such as anger, sadness, happiness, and depression, differ from each other in terms of the extent of their influence on other people. Rosenquist et al.~\cite{rosenquist2011social} used a longitudinal statistical model to analyze a social network of 12,067 people from the Framingham Heart Study~\cite{mahmood2014999}, a long-term, ongoing cardiovascular cohort study of residents of the city of Framingham, MA, to determine whether symptoms of depression in a person are associated with their friends, co-workers, siblings, spouses, and neighbors. To assess depressive symptoms, the researchers used the Center for Epidemiological Scale. The results showed that an association can be found in people at up to three degrees of separation---i.e., from the depressive person's friends to their friends and then to their friends. Researchers have also studied variations of these emotions as shown by online users. Fan et al.~\cite{Fan2014-ak} used a multi-emotions classification model pertaining to anger, joy, disgust, and sadness to determine how these emotions correlate on Weibo---a Chinese microblogging website. Using both Pearson correlation and Spearman correlation, they found that different sentiments have different correlations. Their study showed that correlation among users are high for anger and low for sad sentiments.

Burnap et al.~\cite{Burnap2016-iu} not only included a sentiment analysis but also details pertaining to users' prior party support to predict the outcome of the 2015 UK general election. Having trained a model using nearly 14 million tweets and relying on a range from extremely negative to extremely positive to describe users' sentiments, the researchers predicted that the Labour Party would win the election. In a similar study, Vepsalainen et al.~\cite{Vepsalainen2017-wz} examined how Facebook ``Likes'' can be used to predict election outcomes. They collected 2.7 million data-points using Facebook's Graph API and used Absolute Error to measure the accuracy of their predictions. In that study, the authors were surprised to find that ``Likes'' were not a strong indicator of the election outcome. They look into the reasons they may have achieved this result: skewed demographics, uneven activity by the candidates in social media, and noise in the sample.

\section{Methods} 

\subsection{The dataset}

Our dataset is made up of articles discovered through Altmetric's online database.\footnote{\url{https://www.altmetric.com/}} Their database holds information about millions of scholarly articles, research studies, and news about scientific findings published in a variety of languages and disciplines. We filtered the articles we were targeting to only those that had been shared on a public Facebook page one or more times. We further filtered the articles to only those published in 2017. Choosing this year accomplished three goals. (i) Reactions were released by Facebook in February 2016~\cite{Krug2016}, so any articles we looked at had to be published after that time to have meaningful data on this feature. (ii) Whenever a new feature is rolled out, it takes time for users to learn how to use it; Shah \cite{Shah2016} finds that use of reactions increased from 2.4\% of all interactions in April 2016 to 5.8\% by June 2016, and up to 12.8\% of all interactions by June 2018; by the time of our data collection in early 2019, a large enough subset of users were comfortable expressing themselves with the feature to warrant more scholarly attention. (iii) By the time we began our data collection, a sufficient interval of time had passed for articles to be widely shared and reacted to (between 15 and 30 months). 

Altmetric's database provides URLs for shares of its articles onto public Facebook pages. We used these links to query Facebook's Graph API for information about user reactions to the posts. This process is limited to 200 queries per hour, with each individual query retrieving (i) the click-based reactions, (ii) the number of ``Reshares'' that post received, and (ii) any text included with the share for one share of an article onto a public page. Some articles are shared many times to many different pages; for these, we collected the information for each post and summed all of the reactions into a total reaction score. We did not collect information on the comments added to the posts by users, nor on the article ``Reshares.'' The range of article shares in our dataset was between 1 and 362. The median number of shares that articles in our dataset received was 1. The mean number of shares was 2.30 with a standard deviation of 4.57. Clearly, the distribution of article shares is skewed right---a minority of articles received many times the average number of shares.

We collected data on 356,664 shares of 149,747 scientific articles. Most of the collecting was undertaken between March and July 2019. For this study, however, we were interested in exploring how Facebook users are employing click-based reactions; we thus limit the articles we were looking at to only those that had received one or more of the five special reaction. Our final filtered dataset included 33,662 articles shared onto 178,403 public Facebook pages, all of which received a total of 6,418,053 click-based reactions and 2,051,299 ``Reshares.'' 

For each article, our dataset includes: article title, article abstract, article publication date, the number of public Facebook pages the article was shared onto, and the number of click-based reactions of each category. It also records the text, if any, that was included along with the post. There are also several features that record article topics: ``Subjects,'' which includes the subject areas pertinent to each article that were selected by the authors; ``Scopus subjects,'' which is the subject areas that are recorded in each article's entry in the Scopus database; and ``Publisher subjects,'' which records the subject areas of the journals in which the articles are published. Each of these features can contain one, many, or no subject(s) per article.

In our data collection process, we took the utmost care to respect Altmetric's and Facebook's specifications for how and why their data can be accessed and used. We prioritized avoidance of collecting personally identifying information regarding specific social-media users. Our interests were only in the ways people interact as a whole with scholarly content on social media platforms, not in how specific user's beliefs or opinions may influence their behavior. We recognize that identifying information could in some instances be inferred \textit{a posteriori} from some of the data we collect; however, our method of data collection does not target anything that could be used to consistently identify individual users.

\subsection{Feature analysis and transformation}

Before outlining how we transformed and used the features in our dataset to measure emotional valence, diversity, and intensity, we will first describe some of the basic patterns and relationships we found in our data that informed the decisions we made. Finding the appropriate way to assign importance to this feature or that requires knowledge of what each one signifies and how the features interact with one another. Changes to features must be made carefully and with much deliberation, as important information can be lost or disfigured through the transformation process.
\newcommand{\rowskip}{.18cm}
\newcommand{\ra}[1]{\renewcommand{\arraystretch}{#1}}

\begin{table*}\centering
\resizebox{0.75\columnwidth}{!}{%
\ra{0.6}

\begin{tabular}{crrrrrrr}    \toprule
\textbf{Reaction} & \multicolumn{7}{c}{\textbf{Statistics}} \\

\cmidrule{2-8}

 & Mean & Std. Dev. & Min. & 25\% & 50\% & 75\% & Max.  \\\midrule
\textit{Like}  & 171.2 & 1169.1 & 0 & 13 & 37 & 107 & 156,613    \\
\textit{Love}  & 6.5 & 49.3 & 0 & 1 & 1 & 3 & 6,026    \\
\textit{Wow}  & 6.3 & 52.9 & 0 & 0 & 0 & 2 & 4,120     \\
\textit{Laughter}  & 1.3 & 26.9 & 0 & 0 & 0 & 0 & 4,162     \\
\textit{Sad}  & 2.8 & 34.7 & 0 & 0 & 0 & 0 & 2,778    \\
\textit{Anger}  & 2.6 & 48.6 & 0 & 0 & 0 & 0 & 3,978     \\
\textit{Reshares}  & 60.9 & 386.1 & 0 & 3 & 11 & 37 & 38,272      \\\bottomrule

\end{tabular}
}
\caption{Descriptive statistics for the six click-based reactions and ``Reshares.''}
\label{table:descriptive_stats}
\end{table*}

\subsubsection{Feature analysis}
\begin{figure}
  \includegraphics[width=0.98\textwidth]{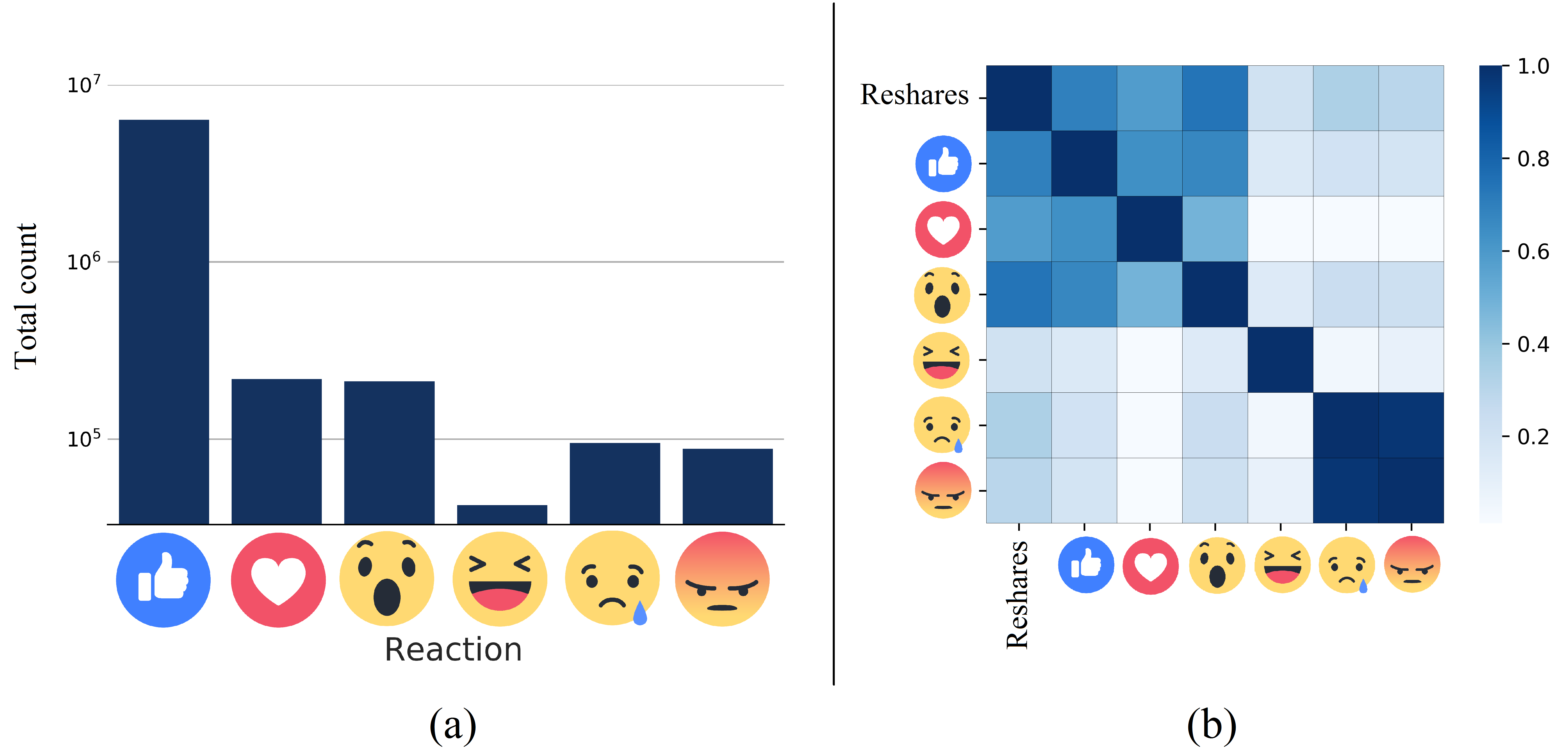}
  \caption{Subfigure (a) shows the distribution of click-based reactions from our dataset; (b) is a heatmap of the Spearman correlation coefficients between click-based reactions to posts and ``Reshares.'' Both visualize values before feature transformation.}
  \label{fig:combined_hist_heatmap}
\end{figure}

Click-based reactions are not evenly distributed throughout the dataset. Figure~\ref{fig:combined_hist_heatmap}a shows the total number of each reaction type in our dataset. We see that there are more ``Likes'' than any other reaction by an order of magnitude. Among the five special reactions, ``Love'' and ``Wow'' are prevalent while negatively valenced reactions such as ``Sad'' and ``Anger'' are less common. Likewise, Table~\ref{table:descriptive_stats} shows descriptive statistics on the six click-based reactions and ``Reshares.'' We might hypothesize that Facebook users are more likely to react positively toward scientific content (or to content on the platform in general), or that positively valenced scientific content is more likely to be propagated through the platform. We might also think that these positive reactions are more common because they are more physically accessible to the user than negatively valenced reactions. While these hypotheses may well carry some weight, we can learn more through closer scrutiny of reaction usage. There are three main factors we will consider when looking at Figure~\ref{fig:combined_hist_heatmap}a: (i) the historical timeline in which Facebook released reactions, (ii) the layout of the Facebook user interface, and (iii) the closeness in semantic meaning of the terms ``Like,'' ``Love,'' and ``Wow.''

First, ``Like'' was the original reaction provided by Facebook. Between 2004 and 2016, ``Like'' and ``Reshare'' were the only click-based reactions available on the platform. By the time the five special reactions were released in 2016, users were well accustomed to using ``Likes'' to respond to a variety of content (for example, positively and negatively valenced posts). It follows that even with the more extensive reaction palette, users were still more likely to employ ``Likes'' out of habit. Second, the five special reactions are reached in the user interface by hovering over the ``Like'' button to open the special reaction palette. ``Sad'' and ``Anger'' are positioned on the far right side of the palette and therefore take the most effort and intention for the user to select, while the ``Like,'' ``Love,'' ``Laughter,'' and ``Wow'' reactions are grouped on the left side of the palette. The ease with which a user can click ``Like'' and that reaction's spatial association with ``Love'' and ``Wow'' might account for these reactions' prevalence in our dataset. This argument is cogent, but can only be extended so far. 

Finally, the term ``Like'' is more semantically related to ``Love'' and ``Wow'' (as in amazement or awe) than any other of the reactions. Further, the icon that represents a ``Like'' is a thumbs-up, a gesture of support and agreement that generally expresses positive sentiment; its positivity can be more closely tied to ``Love'' (which is represented by a heart) and ``Wow'' (represented by an amazed face). These associations account for at least some of the use of this feature. 

We should be careful, however, not to stretch this semantic correlation too far. For example, we might be surprised to find that friend's post about a deceased relative or personal hardship has received ``Likes''; clearly these responses are not meant to show that his or her friends are happy about the circumstances, but rather that they are expressing something more akin to solidarity or sympathy. This example shows that the use of the ``Like'' reaction is not necessarily tied to its semantic meaning. Table~\ref{table:reaction-proportion}, shows the proportion of articles in our dataset that contain each reaction type. Table~\ref{table:reaction-pairings} shows the proportion of articles in our dataset that received at least one of both reactions in each pairing, provide evidence for this effect. We can tell by comparing the proportions of articles with ``Likes'' and other reactions (row 1 of Table~\ref{table:reaction-pairings}) to the values in Table~\ref{table:reaction-proportion} that ``Likes'' are paired with all other reactions almost any time the five special reactions are used. The correlations in Figure~\ref{fig:combined_hist_heatmap}b indicate that ``Likes'' co-vary with ``Love'' and ``Wow'' reactions, but that does not mean that they are not paired with other reaction types as well. 

\def \hfillx {\hspace*{-\textwidth}/3.5 \hfill}

\begin{table}[t]
    \begin{minipage}{0.35\textwidth}
        \centering
            \begin{tabular}{|l|l|}
            \hline
                           & \textbf{Proportion} \\ \hline
            \textbf{Like}  & 0.987               \\ \hline
            \textbf{Love}  & 0.757               \\ \hline
            \textbf{Wow}   & 0.487               \\ \hline
            \textbf{Laughter}  & 0.131               \\ \hline
            \textbf{Sad}   & 0.174               \\ \hline
            \textbf{Anger} & 0.116               \\ \hline
            \end{tabular}
            \caption{Proportion of articles that received one or more of each reaction.}
            \label{table:reaction-proportion}
    \end{minipage}
    \hfillx
    \begin{minipage}{0.6\textwidth}
        \centering
            \begin{tabular}{|c|c|c|c|c|c|c|}
            \hline
                           & \textbf{Like} & \textbf{Love} & \textbf{Wow} & \textbf{Laughter} & \textbf{Sad} & \textbf{Anger} \\ \hline
            \textbf{Like}  & -             & 0.750         & 0.483        & 0.130         & 0.172        & 0.115          \\ \hline
            \textbf{Love}  &               & -             & 0.310        & 0.102         & 0.097        & 0.067          \\ \hline
            \textbf{Wow}   &               &               & -            & 0.098         & 0.116        & 0.081          \\ \hline
            \textbf{Laughter}  &               &               &              & -             & 0.049        & 0.043          \\ \hline
            \textbf{Sad}   &               &               &              &               & -            & 0.076          \\ \hline
            \textbf{Anger} &               &               &              &               &              & -              \\ \hline
            \end{tabular}
            \caption{Proportion of articles that contain one or more of each reaction per pair of reactions.}
            \label{table:reaction-pairings}
    \end{minipage}
\end{table}

Since ``Love'' and ``Wow'' share a semantic and physical closeness to the ``Like'' reaction, we would expect not only to see that they also are used more frequently, but also that the usage of these features can be shown to have positive correlation. Figure~\ref{fig:combined_hist_heatmap}b displays the Spearman correlation coefficients between features. There is a high positive correlation between ``Like,'' ``Love,'' and ``Wow.'' Our intuition that where we see ``Likes'' increase, we also expect ``Love'' or ``Wow'' reactions to increase also is supported by our data. The negatively valenced reactions ``Sad'' and ``Anger'' are also highly correlated with each other. The high positive correlation between ``Reshares'' and ``Like,'' ``Love,'' and ``Wow'' reactions leads us to believe that positive content is more widely shared and reacted to on Facebook---a finding that goes against the conclusion of studies such as~\cite{Fan2014-ak}, which showed that negative emotions lead to greater interaction and dispersion across social networks.

For precisely these reasons, we should not undervalue the appearance of less common reactions such as ``Anger'' or ``Sad.'' Their appearance on an article represents more intentionality and effort on the part of a user to provide a specific response. Because positive reactions are the expected mode of response, and because negative reactions take more effort for user to apply, we decided to weight the different kinds of reactions by the inverse proportion of our expectation of seeing a reaction of that type.

\subsubsection{Feature transformation}

Articles in our dataset are not equally reacted to. A share of an article on one page may result in thousands of responses, while sharing that same article onto another page may result in no reactions at all. The number of reactions may be a signal that a post or article evokes a strong response from users, but it is difficult to account for the context, such as the number of people who saw the post. How many users follow public pages varies from page to page and so the visibility of a post on each page will also vary. It is also difficult to account for Facebook's algorithms that propagate posts into users' news feeds. 

Our method of weighting the click-based reactions is based on the probability with which we would expect to find them on any given article. Weights were determined using a method that is related to Term Frequency-Inverse Document Frequency (TF-IDF)---a method of setting the relative importance of terms within a body of documents that is used in many information-retrieval and recommendation systems today. We refer to the weighting procedure we used as Reaction Frequency-Inverse Document Frequency (RF-IDF).

We changed the raw reaction counts for each article into a proportion of all the click-based reactions the article received. For example, if an article received 6 ``Likes,'' 1 ``Love,'' and 2 ``Wow'' reactions, we would transform those values into $6/(6+1+2)$ ``Likes,'' $1/9$ ``Love,'' and $2/9$ ``Wow'' reactions. We then logarithmically scaled these values. The result of this transformation gave us the \textit{Reaction Frequencies} ($RF_{d_r}$) (Equation~\ref{eq:rf}), where $d_r$ is the count that a given document $d$ received for a certain reaction $r$, and $R$ is the list of all six click-based reactions. 

\begin{equation}
    \mathrm{RF}_{d_r} = \log{\dfrac{d_{r}}{\sum_{r \in R} d_r}}
    \label{eq:rf}
\end{equation}

Next, we needed to reward the rare reactions and penalize the common reactions and to determine the appropriate weights for each reaction type. To do this, we found the probability that a given kind of reaction will be applied to a random article, which is found by taking the number of articles in our dataset that had that reaction type and dividing that number by the number of articles in our dataset. The Inverse Document Frequency (IDF) is the natural logarithm of the inversion of this probability. This value gave us the IDF for each reaction type (Equation~\ref{eq:idf}), where $|D|$ is the number of articles in the dataset and $|D_r|$ is the number of articles in the dataset that received certain reaction $r$. 

\begin{equation}
    \mathrm{IDF}_{r} = \log{\dfrac{\lvert D \rvert}{\lvert D_r \rvert }}
    \label{eq:idf}
\end{equation}

Finally we computed the RF-IDF by multiplying the logarithmically scaled proportion of each reaction type for each article by the IDF for that kind of reaction, as shown in Equation~\ref{eq:rf-idf}.

\begin{equation}
    \mathrm{RF\mbox{-}IDF}_{d_r} = \mathrm{RF}_{d_r} \cdot \mathrm{IDF}_{r}
    \label{eq:rf-idf}
\end{equation}

\subsection{Metrics of emotional diversity and intensity}

With our transformed click-based reactions, we developed metrics to measure the valence, intensity, and diversity of user responses to the articles in our dataset. Valence is the most simple of the three metrics. Since it represents the positive or negative direction toward which a response tends, we had to determine the signals of positive and negative emotions coded in reactions. ``Love'' and ``Anger'' are relatively straight forward, having positive and negative valence respectively. The ``Sad'' reaction could be used to express sympathy or solidarity with a person who has undergone some difficulty, but in the set of posts we are studying it is not likely that users are consistently sharing personal experiences in their posts or research that could inspire this type of sympathetic reaction. Figure~\ref{fig:combined_hist_heatmap} shows a very high correlation coefficient between ``Anger'' and ``Sad''---indeed this pair of features have the highest correlation of all feature pairs. This provides evidence for the idea that these two reactions share a common valence. 

To determine the valence of each article, we thus checked to see whether the value of its ``Love'' reactions was greater than that of its ``Sad'' and ``Anger'' reactions, as shown in Equation~\ref{eq:valence}. 

\begin{equation}
        d_{valence} = 
        \begin{cases}
            -1,& \text{if } (d_{sad} + d_{anger}) > d_{love}\\
            +1,              & \text{else}
    \end{cases}
    \label{eq:valence}
\end{equation}

\noindent
Next we computed the intensity of the response for each article in our dataset. In conceiving our metric, we started with the observation that when providing a click-based reaction to any post on Facebook, each user gets exactly one reaction they can provide, not including ``Reshares'' which can be selected by a user who has ``Liked'' or provided one of the five special reactions. By choosing to select one of the special reactions over the default ``Like'' a user is demonstrating a desire for a more specific response. We interpret this intention and effort as a sign of a stronger emotional reaction. With this observation in mind, we considered intensity as the ratio of the five special reactions to the sum of the six click-based reactions. We began by summing all the reactions for a given document $d$, as shown in Equation~\ref{eq:tot-reacts}.

\begin{equation}
    d_{total\_reacts} = d_{like} + d_{love} + d_{wow} + d_{laughter} + d_{sad} + d_{anger}
    \label{eq:tot-reacts}
\end{equation}

\noindent We then summed up the five special reactions and divided them by the total click-based reactions, as shown in Equation~\ref{eq:intensity}. This bounded our intensity metric between [0,1], where 0 represents an article that received ``Like'' reactions but nothing else, and 1 being the score of an article that received only the special reactions; bounding our metric in this way facilitated comparison between articles. We designed our intensity metric to be sensitive to posts that receive a low reaction count. It was important to us that it be able to identify posts that receive strong signals of emotional intensity without relying on the sheer quantity of reactions.

\begin{equation}
    d_{intensity} = \dfrac{d_{love} + d_{wow} + d_{laughter} + d_{sad} + d_{anger}}{d_{total\_reacts}}
    \label{eq:intensity}
\end{equation}

Finally, we developed a metric to measure the diversity of user responses to each article. Diversity measures how many of the different special reaction types are present for a given article as well as how evenly distributed those reactions are. We disregarded the ``Like'' reaction for our diversity measure. As discussed above, this reaction is quite flexible in use and relies on context for meaning. We use Jensen-Shannon distance (JSD), which uses entropy and Kullback-Leibler divergence to measure the difference between two probability distributions, as the basis for our metric. JSD is found by taking the square root of the Jensen-Shannon divergence score. We preferred JS distance over JS divergence because the former is a true metric of distance and has been shown to satisfy triangle inequality. This latter property improves our ability to compare the results for multiple articles. We preferred JSD over Kullback-Leibler divergence because the latter measure has no upper bound, making comparison between different observations of the measurement difficult. 

We took as a given that the highest diversity of reaction types that could be observed is a uniform spread of types where each reaction is present in equal proportion, and recorded the JSD between this uniform and our observed distribution. JSD is a (0,1) bounded value, where (in our case) 0 indicates that the observed distribution is uniform and 1 shows that an observed distribution varies greatly from uniform (i.e., has only one type of special reaction). 

\begin{equation}
    JSD(\infdiv{P}{Q}) = \sqrt{\frac{1}{2}KLD(\infdiv{P}{M}) + \frac{1}{2}KLD(\infdiv{Q}{M})}
    \label{eq:JSD}
\end{equation}

\noindent
where $P$ and $Q$ are two distributions, $M = \frac{1}{2}(P + Q)$, and $KLD(\infdiv{P}{Q})$ is Kullback-Leibler divergence. Since a uniform distribution, which represents the most diversity an article can receive, will produce 0, we took the complement value:

\begin{equation}
    d_{diversity} = 1 - JSD(\infdiv{\theta_d}{\mathcal{U}\{0,1\}})
    \label{eq:diversity}
\end{equation}

\noindent 
where $\theta_d$ is the distribution of the five special reactions to a given article $d$ and $\mathcal{U}\{0,1\}$ is the discrete uniform distribution from [0,1].

Our metrics are designed to be combined to allow further comparison across articles. For example, diversity evaluates how many different reactions are present on a given article, but not how many reactions are present, or the proportion of the five special reactions to all click-based reactions. By multiplying the diversity by the intensity as displayed in Equation~\ref{eq:divint_index}, we are able to identify articles that received both an intense response, as well as a response that has many different emotions present. 

\begin{equation}
    {d}_{divint\_index} = d_{diversity} \cdot d_{intensity}
    \label{eq:divint_index}
\end{equation}

\noindent
We can also combine valence and intensity scores to produce a polarity score, which reports both strength and direction of a response (Equation~\ref{eq:polarity}). 

\begin{equation}
    {d}_{polarity} = d_{valence} \cdot d_{intensity}
    \label{eq:polarity}
\end{equation}

\subsection{Training a topic model}
Our interest lies beyond how individuals are responding to specific scientific articles: we want to use our data to gain a better understanding of aggregate emotional responses to areas of science. To do so, we wanted to group papers in as logical a manner as possible. We initially considered using the ``Scopus subject'' tags that were associated with each article in the Altmetric dataset, but a great number of articles were missing this feature. Furthermore, users who respond to a post do not necessarily click the link to open the article and read the content before reacting, as shown in~\cite{Gabielkov:2016:SCG:2964791.2901462}. It follows that users are reacting to the content they directly encounter: the Facebook post. As such, we wanted to find ways to maintain as much of the granular detail of the post texts as possible. Following this line of reasoning, we trained our topic model on the texts included with the article shares of each article. 

We took the following steps in building our LDA topic model. We began by combining each text that accompanied shares of a given article. Each article can be shared many times, therefore our ``texts'' were of widely varying length. LDA topic models are especially powerful tools for handling data of this kind~\cite{blei2003latent}. These posts can also be in a variety of languages, so we kept only texts that were in English. We cleaned the texts, removing hyperlinks, punctuation, email addresses, and hashtags. We then created bi-grams and tri-grams (common groupings of two and three words, respectively) from our tokenized texts. We removed stop words such as ``the'' and ``and,'' as well as lemmatized each word, which involves removing inflected endings to transform each word into its dictionary form (e.g., making plural nouns singular and changing verbs into their infinitive form). Finally, we removed words that appeared in less than fifteen documents and those that were in more than half the documents. The result was the corpus we used to train our model. 

We used the LDA model from Python's Gensim library~\cite{rehurek_lrec}. Selecting the number of topics $t$ that a model should identify can be time-consuming. The number may change depending on the goal a researcher has for a model or by the form of their data. Our goal was interpretability: we wanted to be able to easily grasp what a given topic is about, as well as to be able to distinguish one topic from another with relative ease. We also wanted the number of topics to reflect the number of fields we would expect to see in our dataset. There were about thirty different Scopus subject labels (e.g., physics, biochemistry, computer science) applied to the articles in our dataset. We reasoned that setting $t$ somewhere between 15 and 40 would give an appropriate representation of the fields we expect to see in our set. 

We trained seven LDA models, each with a different value of $t$ across the range: $15 \leq t \leq 50$. We compared our models based on their \textit{topic coherence scores} (CS), which is a measurement that tests the degree of semantic similarity between the most representative words in each topic. A higher CS score usually indicates that a model has better identified distinctive topics. The highest coherence score in our set of models was $CS(t=20) = 0.515$. We then used our model with $t=20$ transform the post texts into a distribution of topics. LDA returns the probability that each topic is present for a given document, so we selected the topic that had the highest probability of being present to represent each article. For example, if document $x$ had the topic distribution of $[t_1=0.7, t_2=0.3]$, we would label it as representing topic $t_1$. Table~\ref{table:topics} shows: (i) the twenty discovered topics, (ii) the number of articles in which each topic was the most representative, and (iii) the ten most representative words in each topic. 

\begin{table*}\centering
\resizebox{\columnwidth}{!}{%
\ra{.6}
\begin{tabular}{c|r|p{13cm}}    \toprule
\textit{\textbf{Topic no.}} & \textit{\textbf{Article count}} & \multicolumn{0}{l}{\textit{\textbf{Top ten words}}} \\\midrule
1  & 1,506 & australia, australian, public, year, government, national, country, policy, would, migraine    \\[\rowskip]
2  & 4,045 & study, increase, show, may, high, find, level, result, low, effect    \\[\rowskip] 
3  & 7,929 & not, people, time, say, make, take, do, may, get, many    \\[\rowskip] 
4  & 613   & health, care, need, school, learn, support, practice, student, provide, help    \\[\rowskip] 
5  & 378   & vitamin, diet, exercise, risk, health, muscle, protein, intake, eat, body    \\[\rowskip] 
6  & 3     & state, power, law, grant, core, centre, fire, press, panel, legal    \\[\rowskip] 
7  & 852   & species, population, human, ancient, new, region, area, find, reveal, modern    \\[\rowskip]  
8  & 125   & vaccine, disease, infection, immune system, bacteria, cause, gut, protect, virus, microbiome    \\[\rowskip] 
9  & 168   & planet, light, space, earth, sun, mass, scientist, hot, star, observation    \\[\rowskip] 
10 & 98    & glyphosate, food, exposure, plant, animal, environmental, soil, crop, chemical, organic    \\[\rowskip] 
11 & 64    & disorder, autism, mental, depression, stress, cognitive, physical, self, behavior, social    \\[\rowskip] 
12 & 8     & death, die, skin, florida, emergency, sea, travel, page, kill, cat    \\[\rowskip] 
13 & 105   & use, alcohol, cannabis, health, drug, risk, product, opioid, fda, harm    \\[\rowskip] 
14 & 318   & ice, water, climate change, energy, global, climate, loss, change, air, growth    \\[\rowskip] 
15 & 31    & child, age, old, young, adult, injury, infant, parent, family, year    \\[\rowskip] 
16 & 17    & woman, man, mother, baby, pregnancy, risk, birth, female, male, girl    \\[\rowskip] 
17 & 931   & use, system, method, technology, datum, model, device, network, process, base    \\[\rowskip] 
18 & 1,660 & patient, pain, treatment, risk, cancer, disease, dose, medical, chronic, medicine    \\[\rowskip] 
19 & 4,417 & new, research, read, article, science, publish, paper, journal, scientist, nature    \\[\rowskip] 
20 & 2,157 & brain, cell, human, dna, function, neuron, mouse, protein, gene, genetic    \\  \bottomrule
 \hline
\end{tabular}
}
\caption{The twenty topics discovered by our LDA model with the number of articles where that topic was the most prominent, as well as the top ten words representing each topic.}
\label{table:topics}
\end{table*}

\subsection{Kolmogorov-Smirnov test}

We were interested in testing whether the observed values of our metrics for a subset of articles belonging to a given topic significantly diverge from the values among the rest of the articles. To accomplish this, we used the two-sample Kolmogorov-Smirnov test (KS test). The KS test is a non-parametric statistical test of two continuous, one-dimensional probability distributions. It makes no assumptions about the normality of either distribution, and tests whether the two distributions were sampled from the same populations (or populations with identical distributions). The KS test gives two values: (1) the KS statistic, which is derived from the largest distance (known as the \textit{supremum}) between the cumulative distributions of the two distributions, and (2) a p-value that indicates the significance of the observed KS statistic. The question this p-value answers can be stated as follows: if we assume that the two samples come from the same population, what is the likelihood of observing a given distance between these distributions? A small p-value ($p<.05$) indicates that the probability of seeing such a difference from samples out of the same population is low, and that we can reject the null hypothesis---which in this case indicates that the two samples come from the same population. We formulate our hypotheses in the following way:
\begin{align*}
    H_0 : P=Q \\ 
    H_1 : P \neq Q
    \label{eq:hypotheses}
\end{align*}

\noindent
where P and Q are the two samples used as input for the KS test, $H_0$ is the null hypothesis, in which the two samples come from the same population, and $H_1$ represents the case in which we reject the null hypothesis.

We set our test up as follows. (i) We chose to perform our tests with two of our metrics: diversity and polarity. (ii) We selected several topics that we hypothesized would have distributions of these metrics that departed significantly from the rest of the articles. We chose topics 1 (government), 8 (vaccines), 16 (gender), and 20 (genetics) to test for significantly diversity scores, and topics 8 (vaccines), 10 (agriculture/environmental science), 13 (drug and alcohol), and 14 (climate change) for polarity scores. (iii) We set the significance level needed to reject the null hypothesis at $\alpha = 0.05$. The KS test produces a two-tailed p-value, therefore we can reject $H_0$ only if $p<0.025$ or $p>0.975$, and fail to reject it otherwise. (iv) For each test, we partitioned our data into two groups: the articles representing the given topic and all other articles. (v) We then performed the KS test with the distributions of each metric on these given samples.

\section{Results and discussion}

\begin{figure}
  \includegraphics[width=0.98\textwidth]{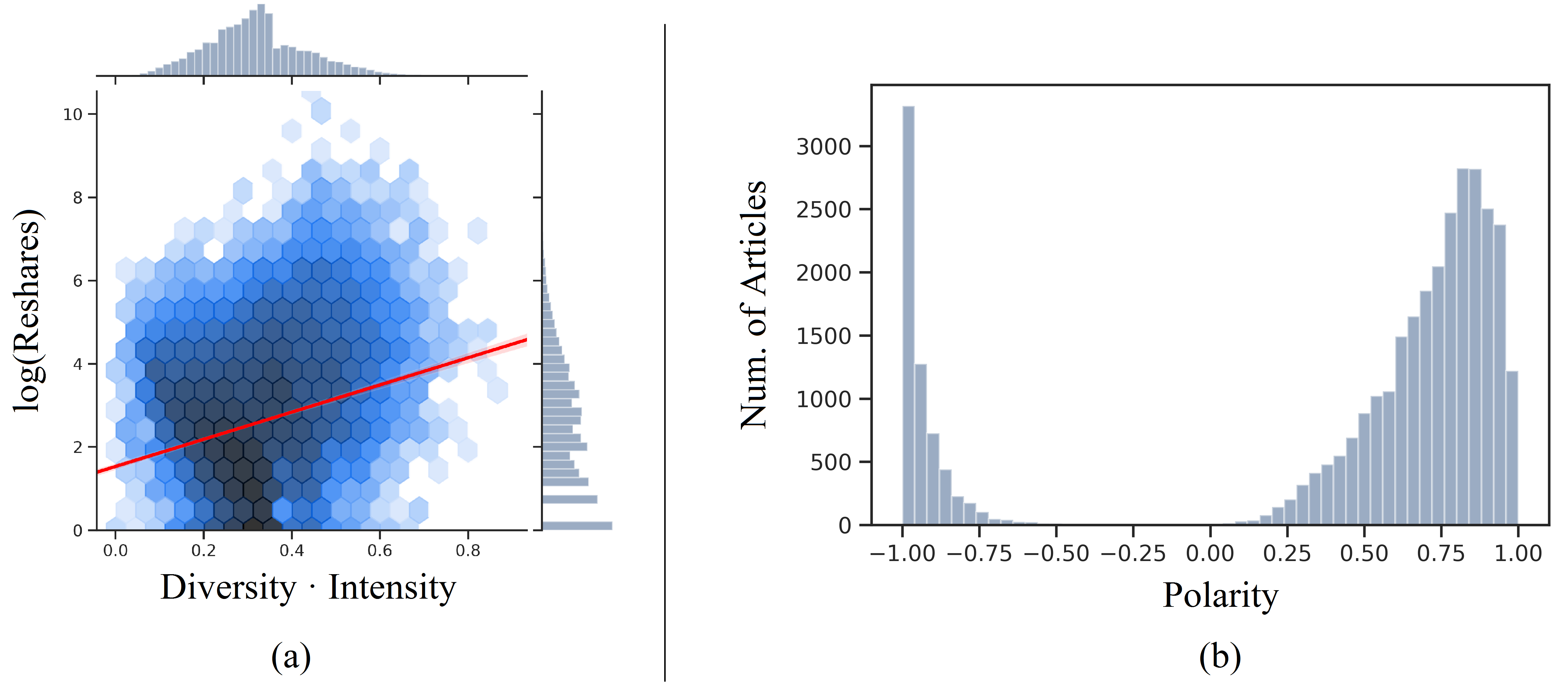}
  \caption{Subfigure (a) is a hexplot showing the distribution of articles in our dataset with regard to the divint index and the logarithmically scaled ``Reshare'' count; a regression line is plotted, showing a positive correlation between the two features; (b) shows the distribution of articles by polarity ($valence \cdot intensity$).}
  \label{fig:hexplot_dist}
\end{figure}

Figure~\ref{fig:hexplot_dist}a displays the distribution of articles along two feature axes: the divint index on the x-axis and the logarithmically scaled ``Reshare'' count on the y-axis. We see that the majority of articles fall within the range of 0.3--0.4 for divint index score ($\mu=0.322$, $med.=0.313$, and $SD=0.117$) and relatively few ``Reshares'' (statistics of this feature are shown in Table~\ref{table:descriptive_stats}). As the divint index of an article increases, the likelihood of it being ``Reshared'' also increases slightly. This relationship is shown with a regression line plotted in red through the figure. 

The distribution of the polarity scores is shown in Figure~\ref{fig:hexplot_dist}b. Though there are more articles that received a positive polarity score, the negative polarity scores are clustered closer to the extreme values. A majority of the articles (80\%) received positive valence, but the most papers in any range are clustered around -1. There are around 3,250 articles that received a polarity score of -1, compared to only 1,250 that received a polarity score of 1. There is a significant drop-off of articles in the range [-0.6, 0]. This behavior should not surprise us: we saw in Figure~\ref{fig:combined_hist_heatmap} that ``Likes'' were not correlated with ``Sad'' or ``Anger'' reactions, but they were highly correlated with ``Love'' reactions. Increased presence of ``Love'' reactions tend to occur with increase ``Like'' reactions, necessarily lowering the intensity score of these cases. On the other hand, increased ``Sad'' or ``Anger'' reactions do not tend accompany increased ``Like'' reactions, pushing the intensity score higher.

\begin{table*}\centering
\resizebox{0.6\columnwidth}{!}{%
\ra{0.4}

\begin{tabular}{cccrr}    \toprule
\multicolumn{2}{c}{\textbf{Topic}} & \phantom{abc} & \multicolumn{2}{c}{\textbf{Test results}} \\

\cmidrule{1-2} \cmidrule{4-5}

Topic No. & Keywords & \phantom{abc} & KS statistic & p-value   \\\midrule
\multicolumn{5}{c}{\textbf{Diversity}}\\
1  & government & & 0.024 & 0.4     \\
8  & vaccines & & 0.077 & 0.429      \\
16 & gender & & 0.107 & 0.984      \\
20 & genetics & & 0.063 & 0.000    \\

\midrule

\multicolumn{5}{c}{\textbf{Polarity}}\\
8  & vaccines & & 0.068 & 0.596      \\
10 & agr./env. science & & 0.151 & 0.021      \\
13  & drugs \& alcohol & & 0.085 & 0.421     \\
14  & climate change & & 0.068 & 0.106     \\ \bottomrule

\end{tabular}
}
\caption{The results of our two-sample KS tests performed for diversity and polarity scores of four topics each.}
\label{table:ks_test}
\end{table*}

Our finding that Facebook users generally react positively to research is in accord with other studies that find that people are more likely to share positive than negative content~\cite{Bazarova:2015:SSE:2675133.2675297}; but our results also indicate that people tend to react with stronger intensity when they have a negative reaction, as seen by the relative lack of articles in the moderately negative range (-0.6, 0] and the large number of papers with a score of -1. Besley and Nisbet~\cite{besley_nisbet} find that scientists view the public as ``emotional'' (as opposed to ``rational'') and ``fear-prone'' when it comes to accepting scientific findings. The prominence of positive emotional reactions expressed toward the articles in our dataset suggests that the ``emotional'' response to science is on average supportive. Because negative reactions or criticism are more salient to a given person~\cite{Lench2011-tt, Bower2001-zq}, it is plausible that these negative reactions stand out more for scientists---especially in cases in which they are more intense. It may be that these negatively valenced but intense reactions disproportionately shape scientists' impressions of the public. 

The results of the KS tests are presented in Table~\ref{table:ks_test}. For diversity scores, topics 16 and 20 showed significant deviations from the rest of the dataset. For polarity scores, only topic 10 showed a deviation significant enough to reject the null hypothesis. The KS test is known to be more likely to result in a failure to reject the null hypothesis when testing with samples of relatively smaller sizes. Topic 16 (gender) only had 17 articles representing the sample, and so we were especially surprised to find such a significant result showing for a sample size so small. 

We were also surprised by cases in which we failed to reject the null hypothesis---especially for either metric with topic 8 (vaccines). Negative emotional reactions to research about vaccines are widely visible and talked about on social-media platforms, but our data indicates that negative reactions on this subject do not significantly deviate from how users respond to research as a whole. It may be the case that because we are tracking scientific articles through Altmetric's database, we are only seeing the responses to vaccine research by those who already accept the scientific findings in that field. To find dissenting opinions we would have to look outside the domain of scientific research.

The distribution of reactions we saw in Figure~\ref{fig:combined_hist_heatmap}a may be the result of our choice of domain for this study. If we were to study, for example, articles in major news outlets it is possible that we would find a higher representation of negative emotions. In targeting shares of scholarly research, we have chosen a domain that is generally considered to be emotionally neutral---though it is not entirely without controversial topics. But even looking at popular news sources, we hypothesize that negative reactions would flag behind positive reactions for the very same reasons (their high correlation with ``Likes'' and the marginal amount of extra work it takes for users to select them). 

Much social-media research is predictive in nature. The models developed by people working in this field often use features from altmetrics to predict what research outcomes will be important, who the rising stars in a field are~\cite{VANDIJK2014R516}, or to discover surprising articles that may have otherwise been overlooked~\cite{Ke7426}. The approach we presented in this paper for feature transformation and generation could also be used in predictive tasks. Predictive models also rely on in-depth knowledge of the data used for training and testing; the analysis we present here could thus stand as the basis for other researchers who are interested in predicting outcomes of science using social-media data. 

We used our metrics to identify aggregate emotional reactions to fields of science, rather than to individual articles, but our approach could be used for individual posts as well. Content managers or platforms might want to find negative or controversial content quickly and efficiently in order to improve user experiences. It could also help scientists and researchers better understand the role they play in shaping the emotional dynamics in broader society. Our metrics provide a way to identify this type of material that is sensitive to sparse reaction profiles so that appropriate responses can be made quickly and efficiently when necessary.

\section{Conclusion and Future Work}

In this paper, we presented a new approach toward the analysis of click-based reactions. We refined this approach through the analysis of a dataset of Facebook reactions to scholarly articles posted on public pages and used it to explore the emotional responses users have to scientific topics. We suggested a method of transforming click-based reactions for analysis that was modeled after the TF-IDF statistic. We drew on concepts from behavioral psychology to develop several metrics for measuring aggregate user behavior with these transformed features. Finally, we used LDA topic modeling and statistical testing to find surprising emotional responses to scientific topics. 

Finding instances of emotional diversity or strong intensity can be beneficial in understanding community dynamics. Conflict and division can be the fault lines on which groups divide, and discovering ways to identify them and make interventions if necessary can improve member cohesion. These lines of division also allow us to appreciate the diversity that exists on many online platforms. Following the contributions of this paper, we plan on using social network analysis to gain better knowledge of how emotions spread within scientific communities. We will investigate the ways emotions affect the dispersion of disinformation and misinformation online, and the role they play in the effective communication of findings to non-experts. Exploring these and other related questions will eventually lead to better outcomes for research and will improve our understanding of the influence emotions have in shaping communication between scientists and the public.

\begin{acks}
This work was partially supported by Argonne National Laboratory under grant number G2A62716.
\end{acks}

\bibliographystyle{unsrtnat}
\bibliography{sample-base}

\begin{thebibliography}{48}
\providecommand{\natexlab}[1]{#1}
\providecommand{\url}[1]{\texttt{#1}}
\expandafter\ifx\csname urlstyle\endcsname\relax
  \providecommand{\doi}[1]{doi: #1}\else
  \providecommand{\doi}{doi: \begingroup \urlstyle{rm}\Url}\fi

\bibitem[Noyes(2019)]{noyes}
Dan Noyes.
\newblock The top 20 valuable facebook statistics, June 2019.
\newblock URL \url{https://zephoria.com/top-15-valuable-facebook-statistics/}.

\bibitem[Costas et~al.(2015)Costas, Zahedi, and
  Wouters]{costas:10.1002/asi.23309}
Rodrigo Costas, Zohreh Zahedi, and Paul Wouters.
\newblock Do ``altmetrics'' correlate with citations? extensive comparison of
  altmetric indicators with citations from a multidisciplinary perspective.
\newblock \emph{Journal of the Association for Information Science and
  Technology}, 66\penalty0 (10):\penalty0 2003--2019, 2015.

\bibitem[Sugimoto et~al.(2017)Sugimoto, Work, Larivi\`{e}re, and
  Haustein]{Sugimoto:10.1002/asi.23833}
Cassidy~R. Sugimoto, Sam Work, Vincent Larivi\`{e}re, and Stefanie Haustein.
\newblock Scholarly use of social media and altmetrics: A review of the
  literature.
\newblock \emph{Journal of the Association for Information Science and
  Technology}, 68\penalty0 (9):\penalty0 2037--2062, 2017.

\bibitem[Thelwall and Nevill(2018)]{THELWALL2018237}
Mike Thelwall and Tamara Nevill.
\newblock Could scientists use altmetric.com scores to predict longer term
  citation counts?
\newblock \emph{Journal of Informetrics}, 12\penalty0 (1):\penalty0 237 -- 248,
  2018.

\bibitem[Didegah et~al.(2018)Didegah, Bowman, and Holmberg]{Didegah2018}
Fereshteh Didegah, Timothy~D. Bowman, and Kim Holmberg.
\newblock On the differences between citations and altmetrics: An investigation
  of factors driving altmetrics versus citations for finnish articles.
\newblock \emph{Journal of the Association for Information Science and
  Technology}, 69\penalty0 (6):\penalty0 832--843, 2018.

\bibitem[Alhoori et~al.(2019)Alhoori, Samaka, Furuta, and
  Fox]{alhoori2019anatomy}
Hamed Alhoori, Mohammed Samaka, Richard Furuta, and Edward~A Fox.
\newblock Anatomy of scholarly information behavior patterns in the wake of
  academic social media platforms.
\newblock \emph{International Journal on Digital Libraries}, 20\penalty0
  (4):\penalty0 369--389, 2019.

\bibitem[Lench et~al.(2011)Lench, Flores, and Bench]{Lench2011-tt}
Heather~C Lench, Sarah~A Flores, and Shane~W Bench.
\newblock Discrete emotions predict changes in cognition, judgment, experience,
  behavior, and physiology: a meta-analysis of experimental emotion
  elicitations.
\newblock \emph{Psychol. Bull.}, 137\penalty0 (5):\penalty0 834--855, September
  2011.

\bibitem[Fredrickson(2001)]{Fredrickson2001-ux}
B~L Fredrickson.
\newblock The role of positive emotions in positive psychology. the
  broaden-and-build theory of positive emotions.
\newblock \emph{Am. Psychol.}, 56\penalty0 (3):\penalty0 218--226, March 2001.

\bibitem[Verduyn et~al.(2013)Verduyn, Van~Mechelen, Tuerlinckx, and
  Scherer]{Verduyn2013-sx}
Philippe Verduyn, Iven Van~Mechelen, Francis Tuerlinckx, and Klaus Scherer.
\newblock The relation between appraised mismatch and the duration of negative
  emotions: Evidence for universality.
\newblock \emph{Eur. J. Pers.}, 27\penalty0 (5):\penalty0 481--494, 2013.

\bibitem[Compton(2003)]{Compton2003-oz}
Rebecca~J Compton.
\newblock The interface between emotion and attention: A review of evidence
  from psychology and neuroscience.
\newblock \emph{Behav. Cogn. Neurosci. Rev.}, 2\penalty0 (2):\penalty0
  115--129, June 2003.

\bibitem[Ram et~al.(2011)Ram, Gerstorf, Lindenberger, and Smith]{Ram2011-zb}
Nilam Ram, Denis Gerstorf, Ulman Lindenberger, and Jacqui Smith.
\newblock Developmental change and intraindividual variability: relating
  cognitive aging to cognitive plasticity, cardiovascular lability, and
  emotional diversity.
\newblock \emph{Psychol. Aging}, 26\penalty0 (2):\penalty0 363--371, June 2011.

\bibitem[Bower and Forgas(2001)]{Bower2001-zq}
Gordon~H Bower and Joseph~P Forgas.
\newblock Mood and social memory.
\newblock In Joseph~P Forgas, editor, \emph{Handbook of affect and social
  cognition , (pp}, volume 457, pages 95--120. Mahwah, NJ, US, Lawrence Erlbaum
  Associates Publishers, xviii, 2001.

\bibitem[Verduyn et~al.(2012)Verduyn, Van~Mechelen, and
  Frederix]{Verduyn2012-xx}
Philippe Verduyn, Iven Van~Mechelen, and Evelien Frederix.
\newblock Determinants of the shape of emotion intensity profiles.
\newblock \emph{Cognition \& Emotion}, 26\penalty0 (8):\penalty0 1486--1495,
  2012.

\bibitem[Verduyn et~al.(2009)Verduyn, Delvaux, Van~Coillie, Tuerlinckx, and
  Van~Mechelen]{Verduyn2009-ib}
Philippe Verduyn, Ellen Delvaux, Hermina Van~Coillie, Francis Tuerlinckx, and
  Iven Van~Mechelen.
\newblock Predicting the duration of emotional experience: two experience
  sampling studies.
\newblock \emph{Emotion}, 9\penalty0 (1):\penalty0 83--91, February 2009.

\bibitem[Levine and Edelstein(2009)]{Levine2009-st}
Linda~J Levine and Robin~S Edelstein.
\newblock Emotion and memory narrowing: A review and goal-relevance approach.
\newblock \emph{Cognition \& Emotion}, 23\penalty0 (5):\penalty0 833--875,
  2009.

\bibitem[Ross et~al.(2018)Ross, Potthoff, Majchrzak, Chakraborty, Ben~Lazreg,
  and Stieglitz]{ross2018diffusion}
Bj{\"o}rn Ross, Tobias Potthoff, Tim~A Majchrzak, Narayan~Ranjan Chakraborty,
  Mehdi Ben~Lazreg, and Stefan Stieglitz.
\newblock The diffusion of crisis-related communication on social media: an
  empirical analysis of facebook reactions.
\newblock In \emph{Proceedings of the 51st Hawaii International Conference on
  System Sciences}, 2018.

\bibitem[Gilbert et~al.(1998)Gilbert, Pinel, Wilson, Blumberg, and
  Wheatley]{Gilbert1998-kn}
Daniel~T Gilbert, Elizabeth~C Pinel, Timothy~D Wilson, Stephen~J Blumberg, and
  Thalia~P Wheatley.
\newblock Immune neglect: A source of durability bias in affective forecasting.
\newblock \emph{Journal of Personality and Social Psychology}, 75\penalty0
  (3):\penalty0 617--638, 1998.

\bibitem[Blei et~al.(2003)Blei, Ng, and Jordan]{blei2003latent}
David~M Blei, Andrew~Y Ng, and Michael~I Jordan.
\newblock Latent dirichlet allocation.
\newblock \emph{Journal of machine Learning research}, 3\penalty0
  (Jan):\penalty0 993--1022, 2003.

\bibitem[Hancock et~al.(2008)Hancock, Gee, Ciaccio, and Lin]{Hancock2008}
Jeffrey~T. Hancock, Kailyn Gee, Kevin Ciaccio, and Jennifer Mae-Hwah Lin.
\newblock I'm sad you're sad: Emotional contagion in cmc.
\newblock In \emph{Proceedings of the 2008 ACM Conference on Computer Supported
  Cooperative Work}, CSCW '08, pages 295--298, New York, NY, USA, 2008. ACM.

\bibitem[Pirzadeh and Pfaff(2014)]{Pirzadeh2014}
Afarin Pirzadeh and Mark~S. Pfaff.
\newblock How do you im when you get emotional?
\newblock In \emph{Proceedings of the 18th International Conference on
  Supporting Group Work}, GROUP '14, pages 243--249, New York, NY, USA, 2014.
  ACM.

\bibitem[Gabielkov et~al.(2016)Gabielkov, Ramachandran, Chaintreau, and
  Legout]{Gabielkov:2016:SCG:2964791.2901462}
Maksym Gabielkov, Arthi Ramachandran, Augustin Chaintreau, and Arnaud Legout.
\newblock Social clicks: What and who gets read on twitter?
\newblock \emph{SIGMETRICS Perform. Eval. Rev.}, 44\penalty0 (1):\penalty0
  179--192, June 2016.

\bibitem[Siravuri et~al.(2018)Siravuri, Akella, Bailey, and
  Alhoori]{Siravuri2018}
Harish~Varma Siravuri, Akhil~Pandey Akella, Christian Bailey, and Hamed
  Alhoori.
\newblock Using social media and scholarly text to predict public understanding
  of science.
\newblock In \emph{Proceedings of the 18th ACM/IEEE on Joint Conference on
  Digital Libraries}, JCDL '18, pages 385--386, New York, NY, USA, 2018. ACM.

\bibitem[{Giuntini} et~al.(2019){Giuntini}, {Ruiz}, {Kirchner}, {Passarelli},
  {Dos Reis}, {Campbell}, and {Ueyama}]{Giuntini}
F.~T. {Giuntini}, L.~P. {Ruiz}, L.~D.~F. {Kirchner}, D.~A. {Passarelli}, M.~D.
  J.~D. {Dos Reis}, A.~T. {Campbell}, and J.~{Ueyama}.
\newblock How do i feel? identifying emotional expressions on facebook
  reactions using clustering mechanism.
\newblock \emph{IEEE Access}, 7:\penalty0 53909--53921, 2019.

\bibitem[Graziani et~al.(2019)Graziani, Melacci, and Gori]{Graziani}
Lisa Graziani, Stefano Melacci, and Marco Gori.
\newblock Jointly learning to detect emotions and predict facebook reactions.
\newblock In Igor~V. Tetko, V{\v{e}}ra K{\r{u}}rkov{\'a}, Pavel Karpov, and
  Fabian Theis, editors, \emph{Artificial Neural Networks and Machine Learning
  -- ICANN 2019: Text and Time Series}, pages 185--197, Cham, 2019. Springer
  International Publishing.

\bibitem[{Raad} et~al.(2018){Raad}, {Philipp}, {Patrick}, and
  {Christoph}]{Raad}
B.~T. {Raad}, B.~{Philipp}, H.~{Patrick}, and M.~{Christoph}.
\newblock Aseds: Towards automatic social emotion detection system using
  facebook reactions.
\newblock In \emph{2018 IEEE 20th International Conference on High Performance
  Computing and Communications; IEEE 16th International Conference on Smart
  City; IEEE 4th International Conference on Data Science and Systems
  (HPCC/SmartCity/DSS)}, pages 860--866, June 2018.

\bibitem[Badache and Boughanem(2017)]{Badache}
Ismail Badache and Mohand Boughanem.
\newblock Emotional social signals for search ranking.
\newblock In \emph{Proceedings of the 40th International ACM SIGIR Conference
  on Research and Development in Information Retrieval}, SIGIR '17, pages
  1053--1056, New York, NY, USA, 2017. ACM.

\bibitem[{Freeman} et~al.(2019){Freeman}, {Roy}, {Fattoruso}, and
  {Alhoori}]{shared_feelings}
C.~{Freeman}, M.~K. {Roy}, M.~{Fattoruso}, and H.~{Alhoori}.
\newblock Shared feelings: Understanding facebook reactions to scholarly
  articles.
\newblock In \emph{2019 ACM/IEEE Joint Conference on Digital Libraries (JCDL)},
  pages 301--304, June 2019.

\bibitem[Tian et~al.(2017)Tian, Galery, Dulcinati, Molimpakis, and
  Sun]{Tian2017-tp}
Ye~Tian, Thiago Galery, Giulio Dulcinati, Emilia Molimpakis, and Chao Sun.
\newblock Facebook sentiment: Reactions and emojis.
\newblock In \emph{Proceedings of the Fifth International Workshop on Natural
  Language Processing for Social Media}, pages 11--16, 2017.

\bibitem[Krebs et~al.(2017)Krebs, Lubascher, Moers, Schaap, and
  Spanakis]{Krebs2017}
Florian Krebs, Bruno Lubascher, Tobias Moers, Pieter Schaap, and Gerasimos
  Spanakis.
\newblock Social emotion mining techniques for facebook posts reaction
  prediction.
\newblock \emph{CoRR}, abs/1712.03249, 2017.

\bibitem[Basile et~al.(2017)Basile, Caselli, and Nissim]{Basile2017-uq}
Angelo Basile, Tommaso Caselli, and Malvina Nissim, editors.
\newblock \emph{Proceedings of the Fourth Italian Conference on Computational
  Linguistics (CLiC-it 2017), Rome, Italy, December 11-13, 2017}, volume 2006
  of \emph{{CEUR} Workshop Proceedings}, 2017.

\bibitem[Rohde et~al.(2004)Rohde, Reinecke, Pape, and Janneck]{Rohde2004}
Markus Rohde, Leonard Reinecke, Bernd Pape, and Monique Janneck.
\newblock Community-building with web-based systems -- investigating a hybrid
  community of students.
\newblock \emph{Computer Supported Cooperative Work (CSCW)}, 13\penalty0
  (5):\penalty0 471--499, Dec 2004.

\bibitem[Hewitt and Forte(2006)]{hewitt2006crossing}
Anne Hewitt and Andrea Forte.
\newblock Crossing boundaries: Identity management and student/faculty
  relationships on the facebook.
\newblock \emph{Poster presented at CSCW, Banff, Alberta}, pages 1--2, 2006.

\bibitem[Burke and Develin(2016)]{Burke:2016:OMF:2818048.2835199}
Moira Burke and Mike Develin.
\newblock Once more with feeling: Supportive responses to social sharing on
  facebook.
\newblock In \emph{Proceedings of the 19th ACM Conference on Computer-Supported
  Cooperative Work \& Social Computing}, CSCW '16, pages 1462--1474, New York,
  NY, USA, 2016. ACM.

\bibitem[Thagard and Kroon(2006)]{thagard2006emotional}
Paul Thagard and Fred~W Kroon.
\newblock Emotional consensus in group decision making.
\newblock \emph{Mind \& Society}, 5\penalty0 (1):\penalty0 85--104, 2006.

\bibitem[Kumar et~al.(2018)Kumar, Hamilton, Leskovec, and
  Jurafsky]{Kumar:2018:CIC:3178876.3186141}
Srijan Kumar, William~L. Hamilton, Jure Leskovec, and Dan Jurafsky.
\newblock Community interaction and conflict on the web.
\newblock In \emph{Proceedings of the 2018 World Wide Web Conference}, WWW '18,
  pages 933--943, Republic and Canton of Geneva, Switzerland, 2018.
  International World Wide Web Conferences Steering Committee.

\bibitem[Zafarani et~al.(2014)Zafarani, Abbasi, and
  Liu]{zafarani_abbasi_liu_2014}
Reza Zafarani, Mohammad~Ali Abbasi, and Huan Liu.
\newblock \emph{Social Media Mining: An Introduction}.
\newblock Cambridge University Press, 2014.

\bibitem[Rosenquist et~al.(2011)Rosenquist, Fowler, and
  Christakis]{rosenquist2011social}
J~Niels Rosenquist, James~H Fowler, and Nicholas~A Christakis.
\newblock Social network determinants of depression.
\newblock \emph{Molecular psychiatry}, 16\penalty0 (3):\penalty0 273, 2011.

\bibitem[Mahmood et~al.(2014)Mahmood, Levy, Vasan, and Wang]{mahmood2014999}
Syed~S Mahmood, Daniel Levy, Ramachandran~S Vasan, and Thomas~J Wang.
\newblock The framingham heart study and the epidemiology of cardiovascular
  disease: a historical perspective.
\newblock \emph{The Lancet}, 383\penalty0 (9921):\penalty0 999 -- 1008, 2014.

\bibitem[Fan et~al.(2014)Fan, Zhao, Chen, and Xu]{Fan2014-ak}
Rui Fan, Jichang Zhao, Yan Chen, and Ke~Xu.
\newblock Anger is more influential than joy: sentiment correlation in weibo.
\newblock \emph{PLoS One}, 9\penalty0 (10):\penalty0 e110184, October 2014.

\bibitem[Burnap et~al.(2016)Burnap, Gibson, Sloan, Southern, and
  Williams]{Burnap2016-iu}
Pete Burnap, Rachel Gibson, Luke Sloan, Rosalynd Southern, and Matthew
  Williams.
\newblock 140 characters to victory?: Using twitter to predict the {UK} 2015
  general election.
\newblock \emph{Elect. Stud.}, 41:\penalty0 230--233, March 2016.

\bibitem[Veps{\"a}l{\"a}inen et~al.(2017)Veps{\"a}l{\"a}inen, Li, and
  Suomi]{Vepsalainen2017-wz}
Tapio Veps{\"a}l{\"a}inen, Hongxiu Li, and Reima Suomi.
\newblock Facebook likes and public opinion: Predicting the 2015 finnish
  parliamentary elections.
\newblock \emph{Gov. Inf. Q.}, 34\penalty0 (3):\penalty0 524--532, September
  2017.

\bibitem[Krug(2016)]{Krug2016}
Sammy Krug.
\newblock Reactions now available globally, Feb 2016.
\newblock URL
  \url{https://newsroom.fb.com/news/2016/02/reactions-now-available-globally/}.

\bibitem[Shah(2018)]{Shah2016}
Pritam Shah.
\newblock {Facebook's new Reactions are being used more \textendash a lot
  more}, Jun 2018.
\newblock URL \url{https://www.quintly.com/blog/new-facebook-reaction-study}.

\bibitem[{\v R}eh{\r u}{\v r}ek and Sojka(2010)]{rehurek_lrec}
Radim {\v R}eh{\r u}{\v r}ek and Petr Sojka.
\newblock {Software Framework for Topic Modelling with Large Corpora}.
\newblock In \emph{{Proceedings of the LREC 2010 Workshop on New Challenges for
  NLP Frameworks}}, pages 45--50, Valletta, Malta, May 2010. ELRA.

\bibitem[Bazarova et~al.(2015)Bazarova, Choi, Schwanda~Sosik, Cosley, and
  Whitlock]{Bazarova:2015:SSE:2675133.2675297}
Natalya~N. Bazarova, Yoon~Hyung Choi, Victoria Schwanda~Sosik, Dan Cosley, and
  Janis Whitlock.
\newblock Social sharing of emotions on facebook: Channel differences,
  satisfaction, and replies.
\newblock In \emph{Proceedings of the 18th ACM Conference on Computer Supported
  Cooperative Work \& Social Computing}, CSCW '15, pages 154--164, New York,
  NY, USA, 2015. ACM.

\bibitem[Besley and Nisbet(2013)]{besley_nisbet}
John~C. Besley and Matthew Nisbet.
\newblock How scientists view the public, the media and the political process.
\newblock \emph{Public Understanding of Science}, 22\penalty0 (6):\penalty0
  644--659, 2013.

\bibitem[van Dijk et~al.(2014)van Dijk, Manor, and Carey]{VANDIJK2014R516}
David van Dijk, Ohad Manor, and Lucas~B. Carey.
\newblock Publication metrics and success on the academic job market.
\newblock \emph{Current Biology}, 24\penalty0 (11):\penalty0 R516 -- R517,
  2014.

\bibitem[Ke et~al.(2015)Ke, Ferrara, Radicchi, and Flammini]{Ke7426}
Qing Ke, Emilio Ferrara, Filippo Radicchi, and Alessandro Flammini.
\newblock Defining and identifying sleeping beauties in science.
\newblock \emph{Proceedings of the National Academy of Sciences}, 112\penalty0
  (24):\penalty0 7426--7431, 2015.

\end{thebibliography}

\end{document}